\input psfig.sty
\documentclass{aa}

\usepackage{graphicx}

\begin{document}

\title{The Infrared Counterpart of the X-ray Burster KS 1731--260\thanks{Based on observations collected at the European Southern Observatory, La
Silla, Chile,  under   proposal  numbers
59.D-0719,  61.D-0542}~~\thanks{RPM and SC contributed equally to this work}}

\author{R.P. Mignani\inst{1}  \and S.Chaty \inst{2} 
\and I.F. Mirabel\inst{3,4} \and S.Mereghetti \inst{5}}

\offprints{rmignani@eso.org}

\institute{ESO, Karl Schwarzschild Str.  2, D-85748 Garching bei
M\"unchen,  Germany  \and The  Open  University,  Department of
Physics and Astronomy Walton Hall,  Milton
Keynes,  MK7 6AA, United Kingdom  \and Service  d'Astrophysique /  CEA, CE-Saclay,
91191 Gif/Yvette, France \and Instituto de Astronom\'\i a y F\'\i sica
del  Espacio/CONICET. cc5,  1428  Bs As,  Argentina  \and Istituto  di
Astrofisica  Spaziale   e  Fisica  Cosmica  CNR,   Sezione  di  Milano
''G.Occhialini'' , via Bassini 15, I-20133 Milano, ITALY}

\date{Received / Accepted }

\titlerunning{KS 1731--260}

\abstract{We present $JK'$ infrared images of the X-ray transient KS 1731--260,
obtained  in  1997  and in  1998  with  IRAC2b  at the  ESO/MPI  2.2-m
telescope  at La  Silla.   Using  as a  reference  the recent  Chandra
position, we confirm  the identification of the X-ray  source with the
previously proposed counterpart (Wijnands et al.  2001b), for which we
measure $J=17.32 \pm  0.2$ and $K' = 16.36 \pm  0.18$.  The source was
entering a low  X-ray state at the epoch of  our observations, and the
accretion disk  was still dominating the infrared  flux.  Indeed, when
compared with the only published  magnitudes in the $J$-band (Orosz et
al.   2001), obtained with  the source  in quiescence,  our photometry
confirms  the  fading of  the  counterpart  during  the decay  of  the
RXTE/ASM lightcurve.
 }

\maketitle

\section{Introduction.}

KS  1731--260  was  discovered  in  the galactic  plane  region  ($  l=
1.073^{\circ};  b=3.653^{\circ}  $) by  the  TTM imaging  spectrometer
aboard the MIR-KVANT observatory (Sunyaev et al. 1989). Soon after the
discovery, several type  I X-ray bursts were detected  by TTM (Sunyaev
et al.  1990),  thus establishing that the source is  an LMXB with an
accreting neutron star.  A long term monitoring of the source with the
RXTE (Muno et  al.  2000) has revealed a  possible correlation between
the properties of the bursts  and the source spectral state.  Coherent
oscillations  at  524 Hz,  possibly  associated  to  the neutron  star
rotation period,  have been discovered by the  RXTE/PCA during several
X-ray bursts (Smith, Morgan \& Bradt  1997) as well as kHz QPOs in the
persistent emission (Wijnands \&  van der Klis 1997).  The measurement
of the photospheric radius expansion  during a burst by RXTE (Smith et
al.  1997; Muno et al.   2000) allowed to estimate the source distance
(7 kpc).  Recently, a superburst has been discovered by the WFC aboard
BeppoSAX (Kuulkers et al.  2002).   A $\approx$ 38 days periodicity in
the  X-ray flux  was  recently discovered  by  Revnivtsev and  Sunyaev
(2001).   While KS  1731--260 is  a  nearly persistent  source in  soft
X-rays (Barret et  al. 1998), it is highly variable  in the hard X-ray
domain, where it  exhibits flux variations occuring on  time scales of
days to weeks  (see, e.g., Kuulkers et al.   2002). However, the X-ray
activity of the source underwent a gradual decrease and since one year
it has entered into a quiescent state (Wijnands et al.  2001a;Wijnands
et  al.  2002).  \\ The  identification of  the companion  star  to KS
1731--260 has been for a long  time hampered both by the high reddening
along  the  galactic  plane and  by  the  lack  of an  accurate  X-ray
position.   The   two  candidate  optical   counterparts  proposed  by
Cherepashuk   et  al.    (1994)  were   excluded  by   later  infrared
observations (Barret  et al. 1998).  Only recently,  the precise X-ray
position provided by Chandra observations (Revnivtsev \& Sunyaev 2002)
has  allowed  to identify  an  optical  counterpart  in the  $I$  band
(Wijnands  et  al.   2001b),   although  without  providing  its  flux
measurement.  Soon after, the counterpart has been observed in the $J$
band  by Orosz et  al.  (2001).  \\ Here,  we present  and independent
confirmation  of the source  identification based  on the  analysis of
images of the KS 1731--260 field obtained back in 1997 and in 1998 with
the IRAC2b  camera at the ESO/MPI  2.2m telescope in the  $J$ and $K'$
bands. The observations are described  in Section 2 and the results in
Section 3.

\section{Observations and Data Reduction.}

The observations of the KS 1731--260 field were carried out in two runs
on  July 19th  1997 and  July 6th/7th  1998 at  the  European Southern
Observatory  (ESO)  in  La  Silla  (Chile)  using  the  ESO/MPI  2.2~m
telescope.  Images of the field were acquired in the infrared with the
IRAC2b camera through the J ($\lambda=1.247~\mu$\rm m, $\Delta
\lambda= 0.290~\mu$\rm m) and K'($\lambda=2.2~
\mu$\rm m, $\Delta \lambda= 0.32~ \mu$\rm m) passbands.  The IRAC2b camera was
mounted at the F/35 infrared adapter of the telescope.  The instrument
was  a Rockwell 256$\times$256  pixels Hg:Cd:Te  NICMOS 3  large format
infrared array  detector.  It was used  with the lens  C, providing an
image scale  of $0\farcs49$/pixel and a field  of $136 \times
136 \mbox{ arcsec}^{2}$.  The observing strategy was identical for the
two observing  runs, with  the source being  observed in both  $J$ and
$K'$ through  a sequence of repeated  exposures of 1  min each.  After
each image of the target field, an image offset by 30 arcsec was taken
to allow for sky subtraction.  The total integration time per passband
was 15 min for the 1997 run,  while two integrations of 9 min each per
passband were collected in 1998 to monitor for short term variability.
The log  of the  observations is summarized  in Table 1.   The typical
seeing conditions for the observations were varying between $0\farcs7$ and $1\farcs2$. \\ After sky subtraction,
the  images  have been  corrected  for  the  instrumental effects,  by
removal  of the bias,  dark current  and flatfielding,  using standard
procedures  available  in  IRAF.   The  final images  have  then  been
registered and  combined. Photometric calibration  was performed using
observations  of  7 standard  stars,  with  some  of them  repeatedly
observed  during   the  night   to  allow  for   accurate  zero-points
determination.

\begin{table}[h]
\begin{center}
\begin{tabular}{|c|c|c|c|}
\hline
 Date &   J  & K' & seeing \\
\hline
 1997.07.19 & 15x1m & 15x1m & $1\farcs2$ \\ 1998.07.06 & 9x1m & 9x1m &
$0\farcs8$ \\ 1998.07.07 & 9x1m & 9x1m & $1\farcs0$ \\
\hline
\end{tabular}
\end{center}
\caption[]{\label {} Observations of KS 1731--260 obtained with the IRAC2b camera at the ESO/MPI 2.2m telescope. The observing dates, the exposure times, the number of exposures per passband and the 
seeing conditions are listed.}
\end{table}

\begin{figure}[h]
\centerline{\hbox{\psfig{figure=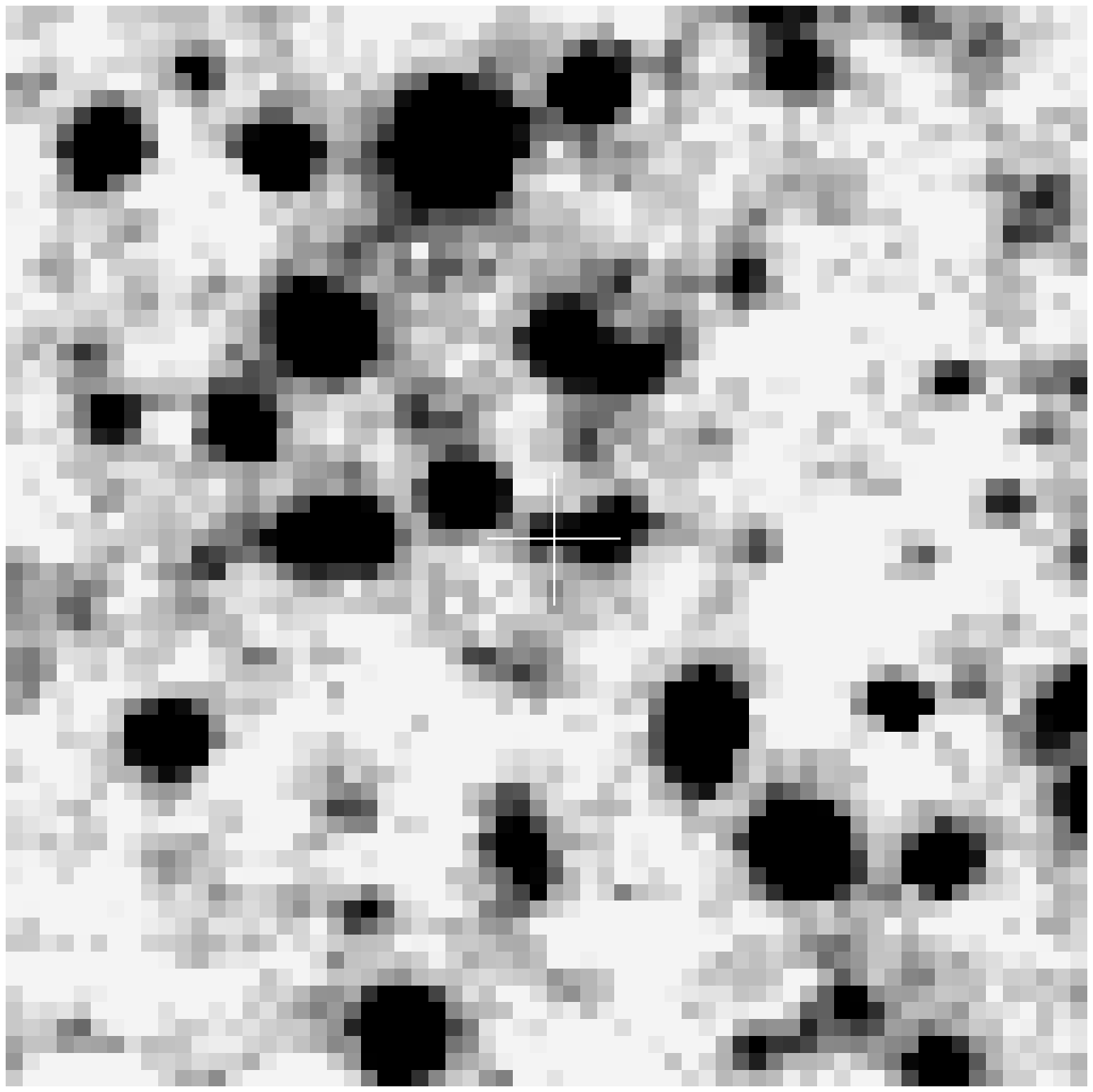,height=8cm,clip=}}}
\vspace{0.5cm}
\centerline{\hbox{\psfig{figure=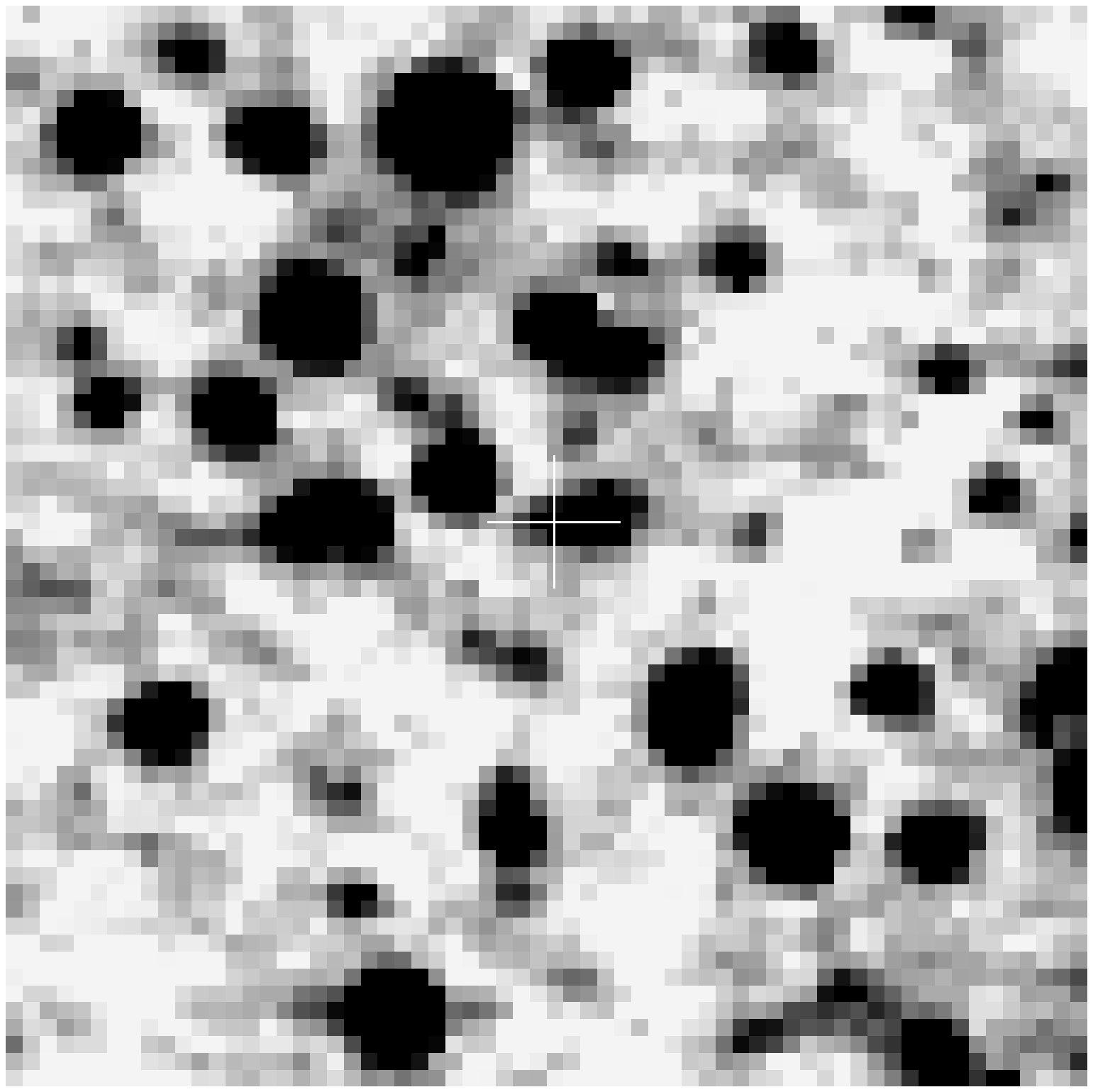,height=8cm,clip=}}}

\caption{({\em top}) $30'' \times 30''$ cutout of the combined $J$ band image 
of the  KS 1731--260 field  obtained on July  6th 1998 with  the IRAC2b
camera at  the ESO/MPI 2.2m telescope.  North to the top,  East to the
left.  The cross  marks the position of the  X-ray source according to
the Chandra  localization (Revnivtsev \&  Sunyaev 2002).  The  size of
the cross arms  corresponds to 3 times the  overall uncertainty on the
source  position  as  given  from  the combination  of  the  intrinsic
coordinates  accuracy and  our  astrometric solution  (see text).
({\em bottom}) Same, but for the $K'$ band image.}

\end{figure}

In order to register the  accurate Chandra coordinates of the source -
$\alpha(J2000)   =   17^{\rm    h}34^{\rm   m}13\fs34   \pm   0\fs04$;
$\delta(J2000)=-26^\circ 05' 18\farcs 7  \pm 0\farcs 6$ (Revnivtsev \&
Sunyaev  2002)  - on  our  images,  we  needed a  precise  astrometric
solution.  Thus,  the image  astrometry has been  computed using  as a
reference the  positions of stars selected from  the recently released
Guide  Star Catalogue  II (GSC-II),  which has  an  intrinsic absolute
astrometric accuracy  of $\approx 0\farcs3$ per  coordinate (McLean et
al.  2002).  A total of 25  GSC-II objects have been identified in one
of the  averaged $J$-band images and used  as astrometric calibrators.
The pixel coordinates  of the reference stars have  been computed by a
two-dimensional  gaussian fitting  procedure  and transformation  from
pixel to  sky coordinates was  then computed using the  program ASTROM
(Wallace  1992), yielding  an rms  of $\sim$  0\farcs18 in  both Right
Ascension  and  Declination, which  we  assume  representative of  the
accuracy  of our astrometric  solution.  The  final uncertainty  to be
attached  to   the  source  position  is  $\sim 0\farcs65$  in  both
coordinates  and   takes  into  account  the  errors   of  the  source
coordinates  ($0\farcs  6$), the  rms  error  of  our astrometric  fit
($\sim 0 \farcs18$)  and the  propagation  of  the intrinsic  absolute
errors on the GSC-II coordinates ($\sim 0\farcs12$).

\section{Results.}

Figure 1 show $30'' \times 30''$  cutouts of the July 6th 1998 $J$ and
$K'$ images centered around the  computed source position and taken as
a reference  because of  the better seeing  conditions (see  Table 1).
The source position,  marked with a cross, clearly  coincides with the
fainter  of  a  doublet  of  stars, only  partially  resolved  in  our
images. The westernmost of the two is star H of Barret et al.  (1998),
while the fainter star  is unambiguously identified as the counterpart
originally proposed by Wijnands  et al.  (2001).  Thus, our astrometry
provides  a   clear  and   independent  confirmation  of   the  source
identification.  \\ In order to obtain reliable magnitude measurements
for the counterpart,  we have applied a star  subtraction alghoritm to
remove star H as well as  other nearby objects.  This step was applied
by  computing a  model  PSF  using the  IRAF/DAOPHOT  package and  was
iterated to minimize the residuals  after PSF subtraction. 

\begin{figure}[h]
\centerline{\hbox
{\psfig{figure=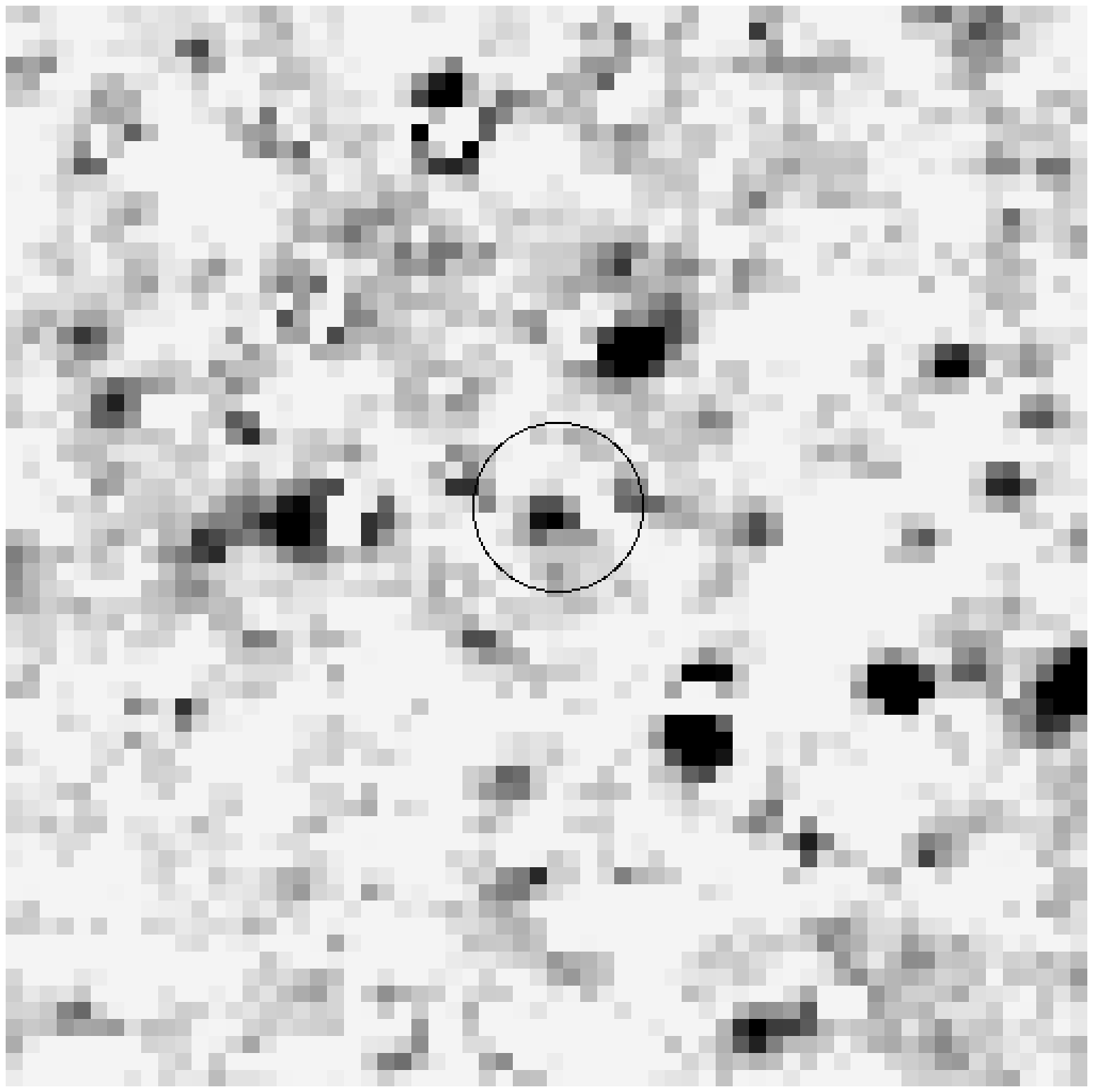,height=4cm,clip=}} 
{\psfig{figure=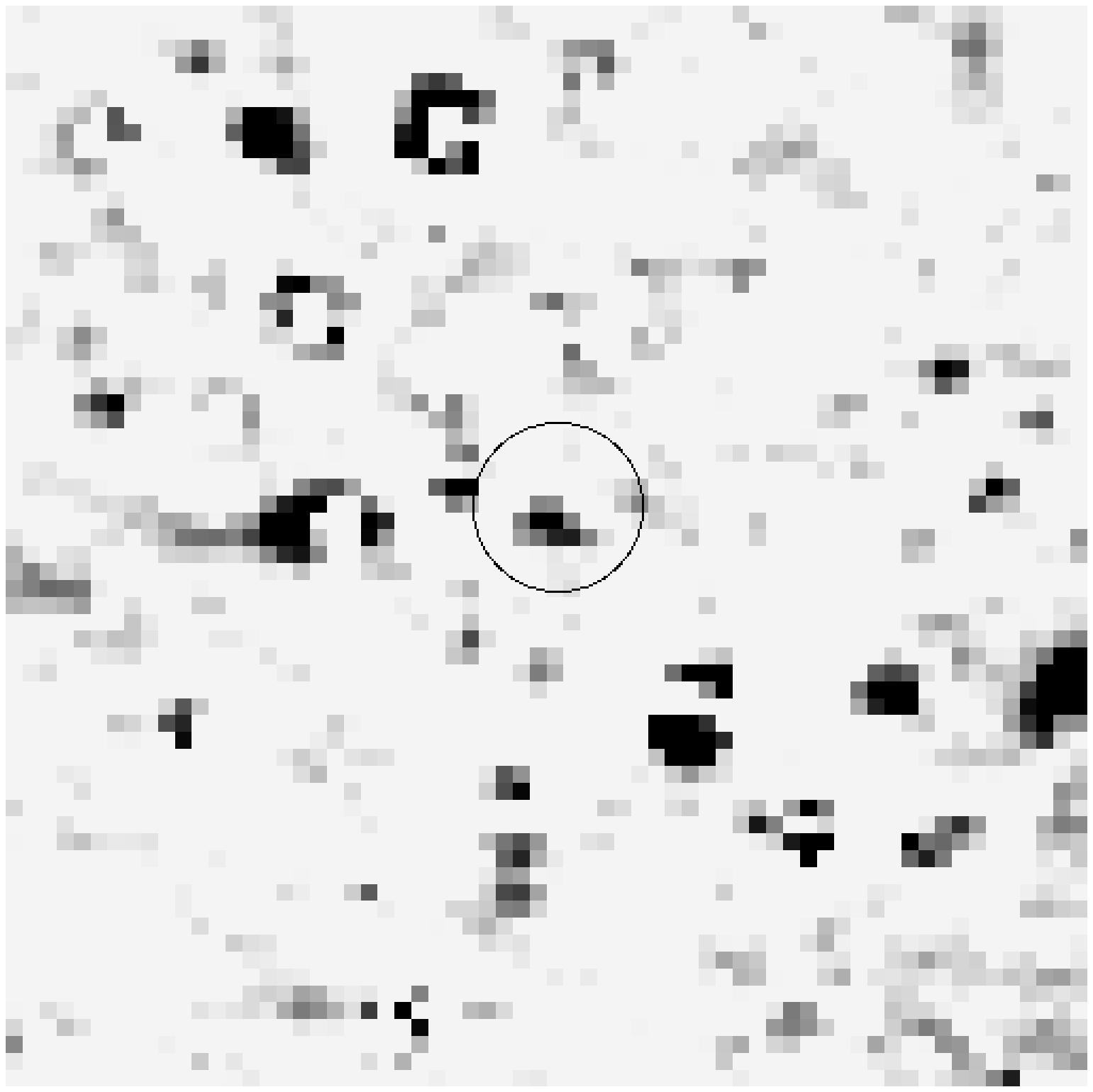,height=4cm,clip=}}
 }
\caption {$15'' \times 15''$ zoom around the source position after
star  subtraction for  the $J$  ({\em  left}) and  $K'$ ({\em  right})
images. The KS 1731--260 counterpart is marked by a circle. }
\end{figure}

\begin{figure}[h]
\centerline{\hbox
{\psfig{figure=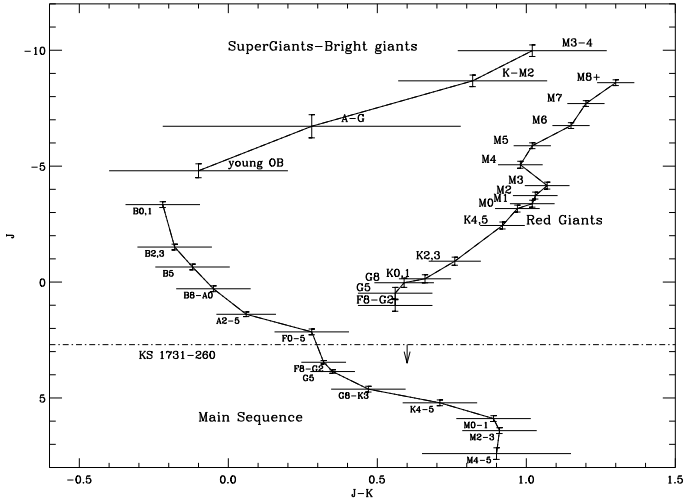,height=7cm,angle=90,clip=}} }
\caption {$J$,$J-K'$ absolute CMD diagram computed for template stars 
(Ruelas-Mayorga  1991).  The position  of  the  companion  star to  KS
1731--260  in  the  diagram  is  anywhere below  the  dot-dashed  line,
corresponding to the absolute magnitude derived from the photometry of
Orosz et al. (2000) assuming a  distance upper limit of $d = 7.0$~ kpc
(Muno et  al. 2001)  and an absorption  $A_{V} \simeq 6$  derived from
Wijnands et al. (2002).  }
\end{figure}

Figure 2
show  a $15''  \times 15''$  zoom  of the  $J$ and  $K'$ band  images,
respectively,  after PSF  subtraction.   The circle  indicates the  KS
1731--260 counterpart.The  source magnitudes  have been computed  to be
$J=17.32 \pm 0.2$  and $K' = 16.36 \pm 0.18$.   To have an independent
assessment  of  our  photometric  calibration  in  the  $J$  band,  we
recomputed the magnitudes of the 13 field objects listed in Table 1 of
Barret  et  al.   (1998),  using  the  IRAF/DAOPHOT  package  for  the
photometry in  crowded fields.  We  found a very good  agreement, with
deviations  within 0.01  magnitudes.   \\ We  have  used our  complete
dataset to  carry out absolute  and relative photometry to  search for
short and long  term variations in the flux of  the counterpart, to be
correlated with the evolution of the X-ray lightcurve.  Unfortunately,
the poorer quality  of the second 1998 observation and  of the 1997 one
did not  allow us to  obtain significant detections and  reliable flux
measurements  for the  counterpart.  However,  compared  with the
value ($J=18.62 \pm  0.21$) measured by Orosz et  al.  (2001) when the
X-ray  source was  in its  low-state, our  $J$ band  magnitude clearly
confirms that the counterpart is indeed variable.

\section{Discussion}

The  comparison  between  the  photometry  of  Orosz  et  al.   (2001)
demonstrates  that, although  the  RXTE/ASM lightcurve  shows that  KS
1731--260  was entering  a low  X-ray state,  the contribution  of the
accretion disc to the observed infrared emission was still dominant at
the time of our 1998  observations.  Thus, the nature of the companion
star  can  not be  constrained  by  our  photometry without  arbitrary
assumption on  the evolution of  the $J-K'$ of the  counterpart during
the transition of the source to the low X-ray state.  However, we note
that if  we take  the $J$-band  magnitude of Orosz  et al.   (2001) as
representative of  the intrinsic flux  of the companion star,  for the
most recent distance estimate of $d  \le 7.0$~ kpc (Muno et al.  2000)
and  the $A_{V}  \simeq  6$ obtained  from  the spectral  fits to  the
combined {\em Chandra}/{\em XMM-Newton} data (Wijnands et al. 2002) we
derive an upper limit on the absolute magnitude $M_{J} \approx 2.7 \pm
0.8$, where  the quoted error  also accounts for the  uncertainties on
the  absorption estimates.  This  value is  plotted in  Fig.3 together
with the theoretical sequences computed from Ruelas-Mayorga (1991). As
it is seen, the diagram suggests  that if the companion is on the main
sequence it must  be of an intermediate spectral  type, probably later
than F. On  the other hand, if the companion has  evolved off the main
sequence,  we  can  likely  exclude it is a Red  Giant.   \\  Infrared
spectroscopy  with the  VLT  will  certainly allow  to  obtain a  more
precise classification of the companion star.

%
%

\begin{acknowledgements}

We thanks  the anonymous  referee for her/his  useful comments  to the
manuscript.   RPM and  SC gratefully  acknowledge the  ESO  Office for
Science, which  funded a visit of  SC at ESO  (Garching), during which
this  work was  finalized.  SC also  acknowledges  support from  grant
F/00-180/A from the  Leverhulme Trust.  The Guide Star  Catalogue - II
is produced by the  Space Telescope Science Institute in collaboration
with the Osservatorio Astronomico  di Torino.  Space Telescope Science
Institute is operated by  the Association of Universities for Research
in Astronomy,  for the  National Aeronautics and  Space Administration
under  contract NAS5-26555.   Additional  support is  provided by  the
Italian   Council  for  Research   in  Astronomy,   European  Southern
Observatory,  Space  Telescope  European  Coordinating  Facility,  the
International   GEMINI   project  and   the   European  Space   Agency
Astrophysics Division.
\end{acknowledgements}


\begin{thebibliography}{}

\bibitem[]{} Barret D., Motch C. \& Predehl P., 1998, A\&A 329, 965 

\bibitem[]{} Kuulkers, E., in 't Zand, J.J.M., van Kerkwijk, M.H., et al., 2002, A\&A 382, 583 


\bibitem[]{} Muno, M.P., Fox, D.W. \& Morgan, E.H., 2000, ApJ 542, 1016

\bibitem[]{} Orosz, J.A., BailYn, C.D., Whitman, K., 2001, ATEL \#75.

\bibitem[]{} Revnivtsev, M.G. \& Sunyaev R.A, 2001, submitted to A\&A, (astro-ph/0108323)

\bibitem[]{} Revnivtsev, M.G. \& Sunyaev R.A, 2002, AstL 28, 19

\bibitem[]{} Ruelas-Mayorga R.A., 1991, Rev. Mex. Astron. Astrof. 22, 27

\bibitem[]{} Smith, D.A., Morgan, E.H. \& Bradt, H., 1997, ApJ 479, L137

\bibitem[]{} Sunyaev R.A. and the Kwant Team, 1989, IAUC 4839 

\bibitem[]{} Sunyaev R.A., Gilfanov, M., Churazov, E., et al., 1990, Sov. Astr. Lett., 16(1), 59 

\bibitem[]{} Wallace, P.T., 1992, Stalink User Note, 5.13

\bibitem[]{} Wijnands, R. \& van der Klis, M., 1997, ApJ 482, L65

\bibitem[]{} Wijnands, R., Groot, P.J., Miller, J.J., Markwardt, C., Lewin, W.H.G. \&  van
der Klis, M., 2001a, ATEL \#72

\bibitem[]{} Wijnands, R., Miller, J.J., Markwardt, C., Lewin, W.H.G. \&  van der Klis, M., 2001b, ApJ 560, L159

\bibitem[]{} Wijnands, R., Guainazzi, M., van der Klis, M. \& Mendez, M., 2002, submitted to
ApJ Lett., (astro-ph/0202398)
                                                                                    
\end{thebibliography}
\end{document}